\documentclass[12pt, fleqn]{article}
\usepackage[mathcal]{euscript}
\usepackage{graphicx}
\usepackage{amssymb}
\usepackage{epstopdf}
\usepackage{latexsym}
\usepackage{indentfirst}
\usepackage{marvosym}
\usepackage{hyperref}

\def\K3{\mathrm K3}

\def\O#1{O({#1})}

\def\double #1{#1{\hbox{\kern-2pt $#1$}}}
\def\p {\partial}

\def\a{{\alpha}}
\def\l{{\lambda}}

\def\g{{\gamma}}
\def\d{{\delta}}

\def\N{{\nabla}}
\def\O{{\Omega}}
\def\half{{1\over 2}}
\def\p{{\partial}}
\def\t{{\theta}}
\def\hat{\widehat}
\def\bar{\overline}

\def\Jb {\bar{J}}
\def\Nb {\bar{N}}

\def\ah{{\widehat\alpha}}

\def\bh{{\widehat\beta}}

\def\o{{\omega}}

\def\L{{\Lambda}}

\def\ll{{\langle}}

\def\half{{1 \over 2}}

\def\p{{\partial}}

\def\pb{{\bar p}}

\def\L{{\Lambda}}

\def\a {{\alpha}}

\def\g {{\gamma}}
\def\d {{\delta}}

\def\ll{{\langle}}

\def\mh{{\widehat m}}
\def\nh{{\widehat n}}

\def\pb{\overline\p}

\def\g{\gamma}

\def\Pib{\overline\Pi}
\def\O{\Omega}
\def\Ob{\overline\O}
\def\N{\nabla}
\def\Nb{\overline\N}
\def\Jb{\overline J}

\def\mh{\widehat m}
\def\nh{\widehat n}

\def\ahh{\widehat a}
\def\bhh{\widehat b}

\def\Ahh{\widehat A}
\def\Bhh{\widehat B}

\def\Mhh{\widehat M}

\newcommand{\be}{\begin{equation}}
\newcommand{\ee}{\end{equation}}
\newcommand{\ba}{\begin{eqnarray}}
\newcommand{\ea}{\end{eqnarray}}
\def\bearst{\begin{eqnarray*}}
\def\eearst{\end{eqnarray*}}

\begin{document}
\setcounter{page}0
\thispagestyle{empty}
\vspace{-70pt}

\begin{flushright}
\makebox[0pt][b]{YITP-SB-09-21}
\end{flushright}

\vspace{38pt}

\center{{\LARGE Compactification of the Heterotic Pure Spinor \\
Superstring I}

\vspace{25pt}

{\large Osvaldo Chand\'ia ${}^\spadesuit$\footnote{email:
ochandia@unab.cl}, William D. Linch, {\sc iii}${}^{
\clubsuit}$\footnote{email: wdlinch3@math.sunysb.edu}\ and Brenno
Carlini Vallilo${}^{\spadesuit}$\footnote{email: vallilo@unab.cl } }

\vspace{15pt}

\center{

${}^{ \spadesuit}${\em Departamento de Ciencias F\'\i sicas,
Universidad Andres Bello \\ Sazie 2315, Santiago, Chile} }

${}^{ \clubsuit}${\em C.N. Yang Institute for Theoretical Physics
and
Department of Mathematics} \\
{\it SUNY, Stony Brook, NY 11794-3840, USA}

\vspace{9pt}

\abstract{ In this paper we begin the study of compactifications of
the pure spinor formalism for superstrings. As a first example of
such a process we study the case of the heterotic string in a
Calabi-Yau background. We explicitly construct a BRST operator
imposing $N=1$ four-dimensional supersymmetry and show that
nilpotence implies K\"ahler and Ricci-flatness conditions. The
massless spectrum is computed using this BRST operator and it agrees with the
expected result.}


\newpage
\setcounter{footnote}0
{{ \parskip=0.0in \tableofcontents }}

\parskip=0.09in

\section{Introduction}

Since the realization that string theory could give rise to
anomaly-free chiral theories, compactifications have been studied in
many different contexts in attempts to make contact with the
observed four-dimensional world. The process of compactification
usually involves breaking of the extended supersymmetry present in
higher dimensional supersymmetric theories.

The compactification procedure for the RNS superstring is well-known
for backgrounds with pure NS fields. If one wants to include RR
fields in the case of Type II string theories, then worldsheet
methods are not available and one is forced to study it using only
supergravity.

In the case of the RNS superstring, compactifications to Calabi-Yau
manifolds and their orbifold limits are standard knowledge in the
field and many interesting physical properties are derived using
worldsheet methods. Alternative descriptions of the RNS superstring
in compactified backgrounds, known as hybrid formalisms, were
developed for two \cite{Berkovits:2001tg}, four
\cite{Berkovits:1994wr}, and six \cite{Berkovits:1994vy} dimensions
by Berkovits and collaborators. Since the roots of the hybrid
formalism are in the RNS superstring, it was not known until
recently \cite{Linch:2006ig}\ how to study compatifications with RR
fields. One of the interesting aspects of the hybrid formalism
approach to RR flux compactifications is that the $N=(2,2)$
superconformal algebra, an essential ingredient in standard CY
compactifications, is still preserved. On the other hand, a drawback
of it is that it is not known how the procedure works if the
starting point of the compactification is not a CY manifold.
Furthermore, computations involving compactification-dependent
states are subtle, and appropriate care should be taken in this
case.\footnote{ One of the authors (BCV) would like to thank Massimo
Bianchi and Pierre Vanhove for pointing out these problems and for
discussions on these issues.}

For these reasons we would like to have another formalism in which
it is possible to study more general flux compactifications. The
pure spinor formalism \cite{purespinors} is the appropriate one.
However, the pure spinor formalism has superspace coordinates
corresponding to all supersymmetries and curved superspaces are not
known explicitly except for maximally symmetric cases and the recent
construction of the full Type IIA superspace for $AdS_4\times\mathbb
CP^3$ \cite{Gomis:2008jt}. For the eleven-dimensional case a
systematic procedure was developed in \cite{Tsimpis:2004gq}.
Although one could use this procedure, four-dimensional
supersymmetry arguments are more effective to attack the present
problem.

Compactifications of the pure spinor formalism is the theme of this
paper. As a first step toward more general backgrounds in heterotic
and type II theories, we will study compactifications of the
heterotic string on a Calabi-Yau 3-fold. The pure spinor formalism
was studied in cases with reduced supersymmetry previously in
\cite{Grassi:2004xc,Berkovits:2005bt,Grassi:2005sb,Chandia:2005fi,
Adam:2006bt}. What is missed by some of these previous works is the
input from the geometry of the Calabi-Yau and the full pure spinor
constraint from ten dimensions. These two ingredients give extra
terms to the BRST charge and these extra terms allow us to derive
on-shell equations for the four-dimensional multiplets.

%

\paragraph{Chiral Superspace and Chiral Coordinates}

As an example of the construction in the next sections let us
consider first the simple case of $N=1$ four-dimensional
supersymmetry. In this case the superspace coordinates are given by
$(x^{\alpha\dot\alpha},\theta^\alpha,\bar\theta^{\dot\alpha})$. The
supersymmetric derivatives are given by \be D_\alpha=\partial_\alpha
+ i\bar\theta^{\dot\alpha}\partial_{\alpha\dot\alpha},\quad \bar
D_{\dot\alpha} =-\partial_{\dot\alpha}
-i\theta^\alpha\partial_{\alpha\dot\alpha}. \ee

It is well-known that a consistent non-trivial constraint on superfields is
$\bar D_{\dot\alpha}\Phi=0$. The easiest way to solve this constraint
is to realize that the chiral variable
$y^{\alpha\dot\alpha}=x^{\alpha\dot\alpha}+i\theta^\alpha\bar\theta^{\dot\alpha}$
is annihilated by $\bar D_{\dot\alpha}$, {\it i.e.} $\bar
D_{\dot\alpha} y^{\beta\dot\beta}=0$. We then construct superfields
depending only on $(y^{\alpha\dot\alpha},\theta^\alpha)$.
Furthermore,  the supersymmetric derivatives and supercharges in
these variables are given by \ba D_\alpha= \partial_\alpha
+2i\bar\theta^{\dot\alpha}\partial_{\alpha\dot\alpha},\quad \bar
D_{\dot\alpha}= \partial_{\dot\alpha},\cr\cr Q_\alpha =
\partial_\alpha , \quad \bar Q_{\dot\alpha} = -\partial_{\dot\alpha}
+2i\theta^\alpha\partial_{\alpha\dot\alpha}, \ea and we have that
$Q_{\alpha} y^{\beta\dot\beta}=0$. This means that any background field
$\Phi(y^{\alpha\dot\alpha})$ is invariant
under chiral supersymmetries.\footnote{Note
that we could also consider dependence on $\bar\theta$ but that is
not a physical superfield, that is, not a representation of the
supersymmetry algebra. } Of course, in Minkowski signature, it is
not possible to consider theories invariant only under the
anti-chiral supersymmetry. (In a Euclidean signature the chiral and
anti-chiral supersymmetries are not related by complex conjugation
and such a symmetry is consistent.) The fact that the
$y^{\alpha\dot\alpha}$ is not real forces us to include its complex
conjugate and one should also consider functions which are not
holomorphic in $y^{\alpha\dot\alpha}$.
This means that chiral coordinates are not useful for reducing
supersymmetry since we cannot have theories constructed only on
subspace parameterized by $y^{\alpha\dot\alpha}$. (Of course the
superpotential is a function on this subspace and one can use
holomorphicity to prove non-renormalization theorems but there is
also the $D$-term.)

It turns out that higher dimensional superspaces also have
chiral-like variables. We will see that after we break
ten-dimensional Lorentz invariance, type I supersymmetry in ten
dimensions will have ``chiral'' variables invariant under
four-dimensional $N=1$ supersymmetry.

\paragraph{Organization} In the next section we introduce the pure
spinor formalism and discuss general concepts that are useful in
later sections. In section 3 we construct curved-space $d$-operators
and the BRST charge for a complex six-dimensional internal manifold
and show that nilpotence and four-dimensional supersymmetry require
that the internal space is a Calabi-Yau manifold. Section 4 contains
a discussion of the spectrum obtained from the cohomology of the
BRST operator constructed in section 3. The final section contains
future directions and open problems.

\section{Preliminary Concepts }

In this section we discuss preliminary material needed for later
sections. We begin with a short review of the pure spinor formalism.
After that we discuss type I supersymmetry preserving only four
dimensional Lorentz symmetry. We close this section with a review of
complex geometry using frames.

\subsection{Review of the Pure Spinor Formalism}

The action of the heterotic string in a flat background is given by

\be \label{flat} S=\int d^2 z [\half \p X^{\mh} \overline\p X_{\mh}
+ p_\ah \overline\p \t^\ah + \o_\ah \pb \l^\ah] + S_\l + S_R,\ee
where $(X^{\mh},\t^\a)$ parameterize the $D=10$, $N=1$ superspace
and $p_\ah$ is the fermionic conjugate momentum. $S_\l$ is the action
for the pure spinor $\l^\ah$ which is defined to satisfy the constraint

\be\label{pure} \l\g^{\mh}\l=0 {\rm ~~for~~} \mh=0 {\rm ~to~} 9.\ee
Although an explicit form  of $S_\l$ in terms of $\l$ and its
conjugate momentum $\o$ requires breaking $SO(9,1)$ (or its
Euclidean version $SO(10)$) to a subgroup, the OPE of $\l^\ah$ with
its Lorentz current $N^{\mh\nh}=\half\o\g^{\mh\nh}\l$ is manifestly
$SO(9,1)$ covariant. The condition (\ref{pure}) implies that $\o$ is
defined only up the gauge invariance

\be\label{gauge} \d\o_\ah=\L^{\mh} (\g_{\mh} \l)_\ah ,\ee for any
$\L^{\mh}$. Finally, $S_R$ is the action for the right-moving
degrees of freedom which describe the reparametrization ghosts and
the heterotic fermions.

It is useful to define the supersymmetric operators in terms of the
free worldsheet fields

\be\label{defdtwo}d_\ah = p_\ah - (\Pi^{\mh} -{1\over
2}\t\g^{\mh}\p\t) (\g_{\mh}\t)_\ah, \quad \Pi^{\mh} = \p X^{\mh} +
\t\g^{\mh}\p\t,\ee which satisfy the OPE's

\be\label{ope}d_\ah(y) d_\bh(z)\to -2 \g^{\mh}_{\ah\bh} \Pi_{\mh}
(y-z)^{-1},\quad d_\ah(y) \Pi^{\mh} (z)\to  (\g^{\mh} \p \t)_\ah
(y-z)^{-1} .\ee The BRST operator and left moving stress energy
tensor are given by

\be\label{brst2}Q = \oint  \l^\ah d_\ah ,\quad
T=-\frac{1}{2}\partial X^{\mh}\partial X_{\mh} -
p_\ah\partial\theta^\ah + T_\lambda \ee where $\l^\ah$ carries
ghost-number $1$. Nilpotency is easily checked using the OPE's
(\ref{ope}) and the pure spinor condition (\ref{pure}). It can be
shown that the cohomological conditions give the equations of motion
and gauge invariances of linearized $N=1, D=10$ supergravity.

In the right moving sector we have the heterotic fermions,
$\bar\Psi_{\mathtt A}$, and the reparametrization ghosts, $(\bar b ,
\bar c)$. The action for them is given by
\begin{equation}
    S_R=\int d^2z [ \bar\Psi_{\mathtt A}\partial \bar\Psi_{\mathtt A}
     + \bar b\partial\bar c] .
\end{equation}
The right moving energy momentum tensor is
\begin{equation}
    \bar T = -\frac{1}{2}\bar\partial X^{\mh}\bar\partial X_{\mh} -
    \bar b \bar\partial \bar c - \bar\partial(\bar b\bar c)
    + \bar T_{\mathbf A},
\end{equation}
where $\bar T_{\mathbf A}$ is the $c=16$ stress energy tensor coming
from the heterotic fermions. Finally, the right moving BRST charge
is given by

\begin{equation}
    \bar Q=\oint ( \bar c\bar T + \bar c\bar\partial\bar c\bar b) .
\end{equation}
Physical vertex operator should be in the cohomology of both $Q$ and
$\bar Q$.

The action in a general curved background can be constructed by
adding the integrated vertex operator to the flat action of
(\ref{flat}) and then covariantizing with respect to the background
super-reparametrization invariance. The result of doing this is
\cite{Berkovits:2001ue}

\be \label{shet} S = \int d^2z \half \Pi^{\ahh} \Pib^{\bhh}
\eta_{\ahh\bhh} + \half \Pi^{\Ahh} \Pib^{\Bhh} B_{\Bhh\Ahh} + d_\ah
\Pib^\ah + \o_\ah \Nb \l^\ah + {\bar\Psi}_{\mathtt A} \N
{\bar\Psi}_{\mathtt A} \ee
$$
+ d_\ah \Jb^I W_I^\ah + \l^\ah \o_\bh \Jb^I U_{I\ah}{}^\bh + S_{FT}
+ S_{bc},
$$
where $\Pi^{\Ahh} = \p Z^{\Mhh} E_{\Mhh}{}^{\Ahh}$ and $\Jb^I = \half
K^I_{{\mathtt A}{\mathtt B}} {\bar\Psi}_{\mathtt A}
{\bar\Psi}_{\mathtt B}$ with the $K$s denoting the generators of the gauge group.
The covariant derivatives are defined as

$$
\Nb\l^\ah = \pb \l^\ah + \l^\bh \Ob_\bh{}^\ah ,\quad \N
{\bar\Psi}_{\mathtt A} = \p {\bar\Psi}_{\mathtt A} + A_I
K^I_{{\mathtt A}{\mathtt B}} {\bar\Psi}_{\mathtt B},$$ where
$\Ob_\bh{}^\ah = \Pib^{\Ahh} \O_{\Ahh\bh}{}^\ah, A_I = \Pi^{\Ahh}
A_{I\Ahh}$ with $\O_{\Ahh\bh}{}^\ah$ being the background connection
for Lorentz and scaling transformations, and $A_{I\Ahh}$ the
connection for background gauge transformations. The
Fradkin-Tseytlin term $S_{FT}$ is given by

\be\label{ft} S_{FT} = \frac{1}{2\pi} \int d^2z r \Phi ,\ee where
$r$ is the world-sheet curvature and $\Phi$ is the dilaton
superfield. Although this term is not necessary for having a
covariant action, it is required to have a quantum conformally
invariant sigma-model action \cite{Chandia:2003hn}
\cite{Bedoya:2006ic}.

\subsection{$N=1$ Ten dimensional Supersymmetry}

\def\il{{\,\mathsf i}}
\def\jl{{\,\mathsf j}}
\def\kl{{\,\mathsf k}}
\def\ll{{\,\mathsf l}}
\def\ml{{\,\mathsf m}}
\def\nl{{\,\mathsf n}}
\def\ic{{\underline\il}}
\def\jc{{\underline\jl}}
\def\kc{{\underline\kl}}
\def\lc{{\underline\ll}}
\def\mc{{\underline\ml}}
\def\nc{{\underline\nl}}

We are interested in a background preserving $N=1$ supersymmetry in
four dimensions. The corresponding supersymmetric derivative algebra
is a sub-algebra of the ten dimensional supersymmetric derivative
algebra. A 16-component, 10-dimensional spinor decomposes into \be
\mathbf{16} \to (\mathbf 2,\mathbf 4)+(\bar\mathbf 2,\bar\mathbf
4)\ee representations of $SL(2,\mathbb C)$ and $SU(4)$. We will
denote the four-dimensional coordinates as $x^a$ or
$x^{\alpha\dot\alpha}$ and the six dimensional coordinates by $y^i$
where the index $i$ goes from $1$ to $6$. To relate vector and
spinor representations of the Lorentz group we use standard sigma
matrices. The six dimensional sigma matrices are $\sigma_i^{IJ}$
where $I,J=1, \dots, 4$ are $SU(4)$ spinor indices and sigma is
antisymmetric in $I$ and $J$. These sigma matrices are related to
the ones with indices down by
\be\label{sigmasix}\bar\sigma^i_{IJ}=\frac{1}{2}\epsilon_{IJKL}\sigma^{i\,
KL}.\ee Other useful identities that the six dimensional sigma
matrices satisfy are \be
\sigma_i^{IJ}\bar\sigma^i_{KL}=\delta^I_K\delta^J_L-\delta^I_L\delta^J_K,\quad
\sigma^{IJ}_i\sigma^{i\, KL}=\epsilon^{IJKL}. \ee

The 16 supersymmetries are now parameterized by complex spinors
$(\eta_\alpha^I,\bar\eta^{\dot\alpha}_I)$ and the
worldsheet spinor variables are now
$(\theta_\alpha^I,\bar\theta^{\dot\alpha}_I)$. The supersymmetry
transformations of the bosonic variables are
\be \delta x^m= i\theta^I\sigma^m\bar\eta_I -i\eta^I\sigma^m\bar\theta_I\ee
\be\delta y^i= i\theta^\alpha\bar\sigma^i\eta_\alpha
-i\bar\eta_{\dot\alpha}\sigma^i\bar\theta^{\dot\alpha},\ee where we
suppressed the index contractions with the sigma matrices. As in four
dimensions it is useful to consider
\be y^{IJ}=y^i\sigma_i^{IJ}, \quad \bar y_{IJ}= y_i\bar\sigma^i_{IJ},\ee
subject to the reality condition
\be(y^{IJ})^\dagger = \frac{1}{2}\epsilon^{IJKL}\bar y_{KL},\ee
inherited from (\ref{sigmasix}).

Since we are interested in preserving only $N=1$ supersymmetry in
four dimensions, we split the $SU(4)$ index to $(\il,\cdot)$ where
$\il=1$ to $3$ and the $\cdot$ denotes a singlet under the $SU(3)$
subgroup of $SU(4)$. Now the odd superspace variables are
$(\theta^\alpha,\bar\theta^{\dot\alpha},\theta_\alpha^\il,
\bar\theta^{\dot\alpha}_\il)$ and the supersymmetry transformations
are given by \be\delta x^m= i\theta\sigma^m\eta
-i\eta\sigma^m\theta+ i\theta^\il\sigma^m\eta_\il
-i\eta^\il\sigma^m\theta_\il,\ee \be\delta y^{\il}=
i\theta^{\alpha\il}\eta_\alpha - i\theta^\alpha\eta_{\alpha}^\il
-i\epsilon^{\il\jl\kl} \bar\eta_{\dot\alpha
\jl}\bar\theta^{\dot\alpha}_{\kl},\ee \be\delta y^{\il\jl}=
i\theta^{\alpha\il}\eta_{\alpha}^\jl
-i\theta^{\alpha\jl}\eta_{\alpha}^\il -i\epsilon^{\il\jl\kl}
\bar\eta^{\dot\alpha}_\kl\bar\theta_{\dot\alpha}
+i\epsilon^{\il\jl\kl}\eta^{\dot\alpha }\theta_{\dot\alpha\kl}.\ee

Note that if we write $y^{\il\jl}$ as $\bar
y^\il=\frac{1}{2}\epsilon^{\il\jl\kl}y^{\jl\kl}$ the reality
condition is just $(y^\il)^\dagger=\bar y^\il$ which means that
$(y^\il,\bar y^\il)$ are usual complex coordinates. In the standard
$SU(4)\to SU(3)\times U(1)$ decomposition, the spinors $\theta^\il$
have $U(1)$ charge $-\frac{1}{2}$ and the singlets $\theta$ have
charge $\frac{3}{2}$ (and the opposite charges for the conjugate
spinors). This is reflected in the supersymmetry transformations
above since $y^\il$ has $+1$ charge.\footnote{Taking care to keep
track of the $U(1)$ charges, we can raise and lower all $SU(3)$
indices at will with the understanding that we only apply Einstein
summation convention when the index carriers have opposite $U(1)$
charges.}

In this notation, the algebra of supersymmetric derivatives in flat
space is given by \ba\label{flatdalgebra}
\begin{array}{lll}
\{ d_\alpha, d_\beta\}=0 & \{ d_\alpha, d_{\dot\alpha}\}
=-2i\partial_{\alpha\dot\alpha} & \{\bar d_{\dot\alpha},
\bar d_{\dot\beta} \} = 0 \\
\{d_\alpha, d_{\beta\il}\}=-2i\varepsilon_{\alpha\beta}\partial_\il
& \{d_\alpha, \bar d_{\dot\alpha\il}\}=0 & \{\bar d_{\dot\alpha},
\bar d_{\dot\beta
\il}\}=-2i\varepsilon_{\dot\alpha\dot\beta}\bar\partial_{\il}\cr
\{d_{\alpha\il},d_{\beta\jl}\}=-4i\varepsilon_{\alpha\beta}
\epsilon_{\il\jl\kl}\bar\partial_\kl ~& \{d_{\alpha\il}, \bar
d_{\dot\alpha\jl}\}=-2i\delta_{\il\jl}\partial_{\alpha\dot\alpha}~&
\{\bar d_{\dot\alpha\il},\bar d_{\dot\beta\jl}\}
=-4i\varepsilon_{\dot\alpha\dot\beta}\epsilon_{\il\jl\kl}\partial_\kl
.
\end{array}
\ea A realization of this algebra in terms of the superspace
coordinates is given by \ba\label{drealization}\begin{array}{l}
d_{\alpha}= \partial_\alpha
+i\bar\theta^{\dot\alpha}\partial_{\alpha\dot\alpha}+
i\theta^\il_\alpha\partial_\il, \cr \bar d_{\dot\alpha} =
-\bar\partial_{\dot\alpha}
-i\theta^\alpha\partial_{\alpha\dot\alpha}
-i\bar\theta^\il_{\dot\alpha}\partial_\il, \cr d_{\alpha\il}=
\partial_{\alpha\il}
+i\bar\theta^{\dot\alpha}_\il\partial_{\alpha\dot\alpha}
-i\theta_\alpha\partial_\il
-2i\epsilon_{\il\jl\kl}\theta^\jl_\alpha\bar\partial_\kl \cr \bar
d_{\dot\alpha\il} = -\partial_{\dot\alpha\il}
-i\theta^\alpha_\il\partial_{\alpha\dot\alpha} + i
\bar\theta_{\dot\alpha}\bar\partial_\il +
2i\epsilon_{\il\jl\kl}\bar\theta^\jl_{\dot\alpha}\partial_\kl
\end{array}\ea

Since we are in flat space, there exist corresponding supercharges
which commute with all these supersymmetric derivatives. However, as
we will not need their full expression here, we will not write them.

The interesting property of the realization (\ref{drealization})
using the notation described earlier is that there exist chiral-like
coordinates analogous to the four-dimensional case described in the
introduction: \be\label{zchiral} z^\il = y^\il
-i\theta^{\alpha\il}\theta_\alpha, \quad \bar z^\il = \bar y^\il
-i\bar\theta^\il_{\dot\alpha}\bar\theta^{\dot\alpha}.\ee These are
invariant under the $SU(3)$ singlet supersymmetries generated by
$(\eta^\alpha,\bar\eta^{\dot\alpha})$ but unlike the
four-dimensional case, we can consistently consider functions of
$(z^\il,\bar z^\il)$ and still have Minkowski signature in
spacetime. Furthermore, when written in these variables the
realization (\ref{drealization}) simplifies to
\ba\label{newrealization}\begin{array}{l} d_{\alpha}=
\partial_\alpha
+i\bar\theta^{\dot\alpha}\partial_{\alpha\dot\alpha}+
2i\theta^\il_\alpha\partial_\il, \cr \bar d_{\dot\alpha} =
-\bar\partial_{\dot\alpha}-i\theta^\alpha\partial_{\alpha\dot\alpha}
-2i\bar\theta^\il_{\dot\alpha}\partial_\il, \cr d_{\alpha\il}=
\partial_{\alpha\il}
+i\bar\theta^{\dot\alpha}_\il\partial_{\alpha\dot\alpha}
-2i\epsilon_{\il\jl\kl}\theta^\jl_\alpha\bar\partial_\kl \cr \bar
d_{\dot\alpha\il} = -\partial_{\dot\alpha\il}
-i\theta^\alpha_\il\partial_{\alpha\dot\alpha} +
2i\epsilon_{\il\jl\kl}\bar\theta^\jl_{\dot\alpha}\partial_\kl,
\end{array}\ea
where now the derivatives $(\partial_\il,\bar\partial_\il)$ are
taken with respect to $(z^\il,\bar z^\il)$. Note that the algebra
(\ref{flatdalgebra}) is preserved. Furthermore, in these new
variables, the corresponding supercharges for the supersymmetries
generated by $(\eta^\alpha,\bar\eta^{\dot\alpha})$ are given by
\be\label{susy1} q_\alpha = \partial_\alpha
-\bar\theta^{\dot\alpha}\partial_{\alpha\dot\alpha},\quad \bar
q_{\dot\alpha}= -\partial_{\dot\alpha} +
i\theta^\alpha\partial_{\alpha\dot\alpha},\ee which means that the
variables $(z^\il,\bar z^\il)$ are invariant under the $SU(3)$ singlet
supersymmetries. Note also that the new ``chiral'' variables
$(z^\il\, , \bar z^\il\, )$ are annihilated by \be\label{subspaces}
\bar d_{\dot\alpha} z^\il\, =0,\quad d_{\alpha\il}\, z^\jl\, =0,\ee
and this is consistent with the algebra (\ref{flatdalgebra}). In
other words, the constraints \be \bar d_{\dot\alpha} \Psi
=d_{\alpha\il}\,\Psi=0\ee on a general superfield $\Psi$ are
integrable.

In what follows, we will assume that our background fields depend
only on these variables. Since the supercharges in (\ref{susy1}) are
independent of $(z^\il,\bar z^\il)$ any background constructed with
them will be invariant under this $N=1$ supersymmetry. This also
means that a background preserving this amount of supersymmetry is
naturally almost-complex. Of course we still have to check that the
background is on-shell. This will be the subject of section 3 where
we will generalize the realization (\ref{newrealization}) to a
curved six-dimensional background.

\subsection{Complex and K\"ahler Geometry Using Frames}

The appropriate language to construct the pure spinor superstring
sigma model in a general background uses frames. Since we want to
study the heterotic string in a Calabi-Yau background, it is useful
to review complex and K\"ahler geometry in this language. The reader
familiar with this material, or willing to accept the
interpretations of the relevant formul\ae{} given in the subsequent
sections, can skip ahead to section 3.
This discussion is based on the definitions and conventions of
\cite{gh}.

A tangent complex index will be denoted by $\il$, as in the previous
subsection, and a coordinate (or ``curved'') index will be denoted
by $\ic$. In a complex manifold of dimension $n$ a hermitian metric
is given in local coordinates by\footnote{A bar over an
antiholomorphic index will not be used unless it is necessary.} \be
ds^2= g_{\ic\jc}dz^\ic\otimes d\bar z^\jc\,. \ee The Riemannian
metric on this manifold is given by ${\rm Re } (ds^2)$ and the
imaginary part of $ds^2$ is given by \be \omega
=ig_{\ic\jc}dz^\ic\wedge d\bar z^\jc, \ee and is called the
associated $(1,1)$-form (or K\"ahler form). An hermitian {\it
coframe} is defined by two matrices $(E^\il{}_\ic,\bar E^\il{}_\jc)$
such that \be\label{coframe} ds^2= g_{\ic\jc}dz^\ic\otimes d\bar
z^\jc=E^\il{}_\ic\,\bar E^\il{}_\jc\, dz^\ic\otimes d\bar z^\jc =
E^\il\otimes\bar E^\il, \ee where $E^\il=E^\il{}_\ic\, dz^\ic$ and
$\bar E^\il=\bar E^\il{}_\ic \,d\bar z^\ic$. Using the coframe, the
associated $(1,1)$-form is given by \be \omega =iE^\il\wedge\bar
E^\il. \ee As usual, the exterior derivative is $d=
\partial + \bar\partial = dz^\ic\partial_\ic + d\bar
z^\ic\bar\partial_\ic$. We can compute the exterior derivative of
the coframe giving \be\label{exterior} d E^\il = (d
E^\il{}_\ic)\wedge dz^\ic = [(\bar\partial E^\il{}_\ic)E^\ic{}_\jl
-\bar E^\ic{}_\il \partial\bar E^\jl{}_\ic]\wedge E^\jl + T^\il, \ee
where $T^\il$ is a $(2,0)$-form defined by the equation above. Its
explicit expression is \be T^\il = (\partial
E^\il{}_\ic)E^\ic{}_\jl\wedge E^\jl + (\partial\bar E^\jl{}_\ic)\bar
E^\ic{}_\il\wedge E^\jl \ee with $E^\ic{}_\il= (E^\il{}_\ic)^{-1}$.
The complex manifold  is  {\it K\"ahler} if $T^\il=0$. Equation
(\ref{exterior}) can be written in the form \be d E^\il =
\Omega^\il{}_\jl\wedge E^\jl +T^\il \ee where
$\Omega^\il{}_\jl=(\bar\partial E^\il{}_\ic) E^\ic{}_\jl -\bar
E^\ic{}_\il \partial\bar E^\jl{}_\ic$ and satisfies $\Omega +
\bar\Omega^\dagger=0$. Such a connection $\Omega$ is compatible with
both the metric and complex structure. To see this more clearly,
note that \be\label{integrability} d(E^\il\wedge\bar E^\il)=
T^\il\wedge\bar E^\il -E^\il\wedge\bar T^\il. \ee From this last
equation we can also see the standard definition of a K\"ahler
manifold, that is, $d \omega=0$ if $T^\il=0$. The equations above
allow us to define covariant exterior derivatives $\nabla$ and
$\bar\nabla$ given by \be \label{covdev} \nabla=\partial + (\bar
E^{-1}\partial\bar E)^\il{}_\jl, \quad \bar\nabla = \bar\partial -
(\bar\partial E E^{-1})^\il{}_\jl. \ee With this definition we can
say that $E^\il$ is covariantly holomorphic \be \bar\nabla E^\il =0,
\ee while the holomorphic covariant exterior derivative $\nabla$
defines the torsion \be \nabla E^\il= T^\il, \ee

In the case of vanishing torsion, these last two equations say
\be\label{structure} \bar\nabla_\ic E^\il{}_\jc=0, \quad\quad
\nabla_\ic E^\il{}_\jc = \nabla_\jc E^\il{}_\ic, \ee where the
second equation translates to the usual $\partial_\ic
g_{\jc\kc}=\partial_\jc g_{\ic\kc}$. One should be careful to note
that the definition of covariant derivatives acts differently on the
frames $E^\il$ and $\bar E^\il$, {\it i.e.} $(\nabla)^\dagger \neq
\bar\nabla$. This is because the connection $\Omega$ defined above
is skew-hermitian: \be \Omega + \Omega^\dagger=0 \to (d +
\Omega)^\dagger = d - \Omega, \ee so the analogous expressions for
the covariant derivatives (\ref{covdev}) for $\bar E^\il$ have
opposite signs and we have, in the case of vanishing $\bar T^\il$,
\be \nabla_\ic\bar E^\il{}_\jc=0, \quad\quad \bar\nabla_\ic\bar
E^\il{}_\jc=\bar\nabla_\jc\bar E^\il{}_\ic .\ee

\paragraph{Curvature}
We can define new covariant derivatives using the inverse of the
coframe matrices. \be \nabla_\il = E^\ic{}_\il \nabla_\ic, \quad
\bar\nabla_\il=\bar E^\ic{}_\il\bar\nabla_\ic. \ee
Also, note that because of (\ref{structure}) we have \be
\bar\nabla_\ic E^\jc{}_\il=0, \quad\quad E^\il{}_\jc\nabla_\ic
E^\kc{}_\il=E^\il{}_\ic\nabla_\jc E^\kc{}_\il .\ee
Using the new
covariant derivatives above we can rewrite these expressions as
\be\label{structure2} \bar\nabla_\jl E^\jc{}_\il=0, \quad\quad
\nabla_\il E^\kc{}_\jl=\nabla_\jl E^\kc{}_\il. \ee
As usual, the curvature is defined from the commutators of covariant
derivatives. Since the manifold is hermitian, we have $[\nabla_\ic\,
,\nabla_\jc\, ]=0$. The same will be true for $\nabla_\il$ precisely
if the second equation in (\ref{structure2}) holds. So, in terms of
the new covariant derivatives the K\"ahler condition is $[\nabla_\il
\, ,\nabla_\jl \, ]=0$. We will see how this condition arises from
nilpotence of the BRST charge in section 3.

The non vanishing part of the curvature matrix can be defined as \be
R_{\ic \bar\jc} = [\nabla_\ic\, ,\bar\nabla_\jc\, ]. \ee Since the
first equation of (\ref{structure2}) holds we have that \be R_{\il
\bar \jl} = E^\ic{}_\il \bar E^\jc{}_\jl R_{\ic
\bar\jc}=[\nabla_\il\, ,\bar\nabla_\jl\, ].\ee
Due to all the symmetries the curvature matrix has when the manifold
is K\"ahler, there are three equivalent ways to write the
Ricci-flatness condition. The first is the usual $\mathrm {Ric}=0$,
the second is ${\rm Tr} (R_{\il \bar \jl}) =0$ and the last one is
$\delta^{\il\jl}R_{\il \bar \jl} =\delta^{\il\jl}[\nabla_\il\,
,\bar\nabla_\jl\, ]=0$. Again, in section 3 we will see how this
last equation appears from nilpotence of the BRST charge.

\paragraph{Vector bundles}
Since we are studying the heterotic string, we know vector bundles
also appear in the theory and couple to the background in a
non-trivial way. Consider that our manifold comes with additional
structure given by a gauge 1-form \be {\mathcal A} = {\mathtt A}_\il
E^\il + \bar{\mathtt A}_\il\bar E^\il = A_\il^\Sigma {\mathbf
T}_\Sigma E^\il + \bar A_\il^\Sigma{\mathbf T}_\Sigma \bar E^\il,
\ee where ${\mathbf T}_\Sigma$ are the gauge algebra generators.
We generalize the covariant derivatives above to include this
gauge 1-form connection \be {\mathcal D}_\il = \nabla_\il - {\mathtt
A}_\il,\quad\quad \bar{\mathcal D}_\il = \bar\nabla_\il -
\bar{\mathtt A}_\il.\ee
Computing again the conditions that give K\"ahler and Ricci-flatness
$ [ {\mathcal D}_\il\, ,{\mathcal D}_\jl\, ]=0 $  and  $
\delta^{\il\jl}[{\mathcal D}_\il\, ,\bar{\mathcal D}_\jl\, ]=0$ they
factorize into original K\"ahler and Ricci-flatness and holomorphic
YM equations, {\it i.e.} $F_{\il\jl}=0$ and $\delta^{\il\jl}
F_{\il\bar\jl}=0$.

\section{An Ansatz for the $d$-operator algebra and BRST Charge}

The expression for the $d$-operators in a general curved background
was derived in \cite{Berkovits:2001ue}. It is given by
\be\label{generald} d_{\hat\alpha} = E_{\hat\alpha}{}^{\hat M} \Big[
P_{\hat M} + \frac{1}{2} B_{\hat M\hat N}(\partial Z^{\hat N}
-\bar\partial Z^{\hat n}) -\Omega_{\hat
M}{}^{\hat\beta}{}_{\hat\gamma}
\lambda^{\hat\gamma}\omega_{\hat\beta} - {\mathtt A}_{\hat M}^\Sigma
{\mathbf J}_\Sigma \Big], \ee where $P_{\hat M}$ are the momenta
conjugate to the worldsheet variables defined as $P_{\hat M}=\delta
S/\delta(\partial_0 Z^{\hat M})$. The nilpotence of the BRST charge
is computed using Poison brackets $[ P_{\hat M},Z^{\hat N} ]_{PB}
=\delta^{\hat N}_{\hat M}$ and
$[\lambda^{\hat\alpha},\omega_{\hat\beta}]_{PB}=\delta^{\hat\alpha}_{\hat\beta}$.
Note that the background field $B_{\hat M\hat N}$ does not mix with
the other background fields in (\ref{generald}) when we compute the
nilpotence condition. This mixing only occurs when computing
holomorphicity of the BRST current. In a flat background the $d$
operator reduces to $d_{\hat\alpha} = E_{\hat\alpha}{}^{\hat M}
P_{\hat M}$ (ignoring the contribution from the flat $B_{\hat M\hat
N}$) and using the expression for the flat frame in the $4+6$
notation we get precisely (\ref{drealization}) after replacing the
conjugate momenta by the corresponding derivatives.
The flat space BRST charge is \be\label{flatq} Q=\oint (
\lambda^\alpha d_\alpha + \bar\lambda^{\dot\alpha}\bar
d_{\dot\alpha} +\lambda^{\alpha \il} d_{\alpha \il} +
\bar\lambda^{\dot\alpha \il} \bar d_{\dot\alpha \il} ),\ee and it
will square to zero if the ghosts satisfy the pure spinor constraint,
reduced to $4+6$ notation \ba \lambda^\alpha\bar\lambda^{\dot\alpha}
+ \lambda^{\alpha \il}
\bar\lambda^{\dot\alpha \il} =0,\\
\lambda^\alpha \lambda^\il_\alpha - \epsilon^{\il\jl\kl}
\bar\lambda^\jl_{\dot\alpha}\bar\lambda^{\dot\alpha \kl}=0,\\
\bar\lambda^{\dot\alpha} \bar\lambda^\il_{\dot\alpha}
- \epsilon^{\il\jl\kl}\lambda^\jl_{\alpha}\bar\lambda^{\alpha \kl}=0.
\ea

We want to generalize this to a flat four-dimensional background
plus a curved six dimensional one. We must find the appropriate
generalization of the $d$ operators for this case. The first thing
to note is that if they are generalized to covariant derivatives
$(\nabla_\alpha,\bar\nabla_{\dot\alpha},\nabla_{\alpha\il},\bar\nabla_{\dot\alpha\il})$
satisfying the following algebra
\begin{eqnarray}\label{algebrasolution}
\hspace{-1.4cm}
\begin{array}{lll}
\{ \nabla_\alpha , \nabla_\beta\} = 0
    &\{\nabla_\alpha, \bar \nabla_{\dot \alpha}\}=-2i \nabla_{\alpha\dot\alpha}
    &\{\bar \nabla_{\dot \alpha} , \bar \nabla_{\dot \beta} \}=0\\
\{ \nabla_\alpha , \nabla_{\beta \il} \} =- 2i \varepsilon_{\alpha \beta} \nabla_\il
    &\{\nabla_\alpha, \bar \nabla_{\dot \alpha\il} \}=0
    &\{\bar \nabla_{\dot \alpha} , \bar \nabla_{\dot \beta \il} \}
    =-2i \varepsilon_{\dot \alpha \dot \beta} \bar \nabla_\il\\
\{ \nabla_{\alpha \il} , \nabla_{\beta \jl}\} = -4i
\varepsilon_{\alpha \beta} \epsilon_{\il\jl\kl} \bar \nabla_\kl
    &\{\nabla_{\alpha \il}, \bar \nabla_{\dot \alpha \jl}\}
    =-2i \delta_{\il\bar\jl} \nabla_{\alpha\dot\alpha}
    &\{\bar \nabla_{\dot \alpha \il} , \bar \nabla_{\dot \beta \jl} \}
    =-4i \varepsilon_{\dot \alpha \dot \beta}\epsilon_{\bar\il\bar\jl\bar\kl}\nabla_\kl
\end{array}
\end{eqnarray}
the BRST charge will be nilpotent. Here, the covariant derivatives
$(\nabla_{\alpha\dot\alpha}, \nabla_\il\, ,\bar\nabla_\il\, )$ are
defined by these equations. Using the variables defined in section
2.2 we can write the spinor covariant derivatives
as \ba\label{curvedrealization}\begin{array}{l} \nabla_{\alpha}=
\partial_\alpha +i\bar\theta^{\dot\alpha}\nabla_{\alpha\dot\alpha}+
2i\theta^\il_\alpha\nabla_\il, \cr \bar \nabla_{\dot\alpha} =
-\bar\partial_{\dot\alpha}-i\theta^\alpha\nabla_{\alpha\dot\alpha}
-2i\bar\theta^\il_{\dot\alpha}\nabla_\il, \cr \nabla_{\alpha\il}=
\partial_{\alpha\il}
+i\bar\theta^{\dot\alpha}_\il\nabla_{\alpha\dot\alpha}
-2i\epsilon_{\il\jl\kl}\theta^\jl_\alpha\bar\nabla_\kl \cr \bar
\nabla_{\dot\alpha\il} = -\partial_{\dot\alpha\il}
-i\theta^\alpha_\il\nabla_{\alpha\dot\alpha} +
2i\epsilon_{\il\jl\kl}\bar\theta^\jl_{\dot\alpha}\nabla_\kl,
\end{array}\ea
The higher order dependence on $\theta$s come from the derivatives
$(\nabla_{\alpha\dot\alpha},\nabla_\il\, ,\bar\nabla_\il\, )$. Note
that the equations (\ref{curvedrealization}) can be put in the form
(\ref{generald}) with the spin connection term $\Omega_{\hat
M}{}^{\hat\beta}{}_{\hat\gamma}
\lambda^{\hat\gamma}\omega_{\hat\beta}$ and the gauge connection
term ${\mathtt A}_{\hat M}^\Sigma {\mathbf J}_\Sigma$ inside the
bosonic covariant derivatives. Since the background does not break
four-dimensional Lorentz symmetry, the covariant derivative
$\nabla_{\alpha\dot\alpha}$ is just $\partial_{\alpha\dot\alpha} +
{\mathcal O}(\theta^2)$ and nothing will depend on
$x^{\alpha\dot\alpha}$. Moreover, since we are imposing that the
background is invariant under the $N=1$ supersymmetry, the background
cannot depend on $(\theta^\alpha,\bar\theta^{\dot\alpha})$. We will
now derive the restrictions imposed by these conditions.

Repeated application of the
Jacobi identities
\begin{eqnarray}
&&\hspace{-1.8cm}
(-)^{AC}[\nabla_A , [\nabla_B, \nabla_C\}\}
+(-)^{BA}[\nabla_B , [\nabla_C, \nabla_A\}\}
+(-)^{CB}[\nabla_C , [\nabla_A, \nabla_B\}\} =0,
\nonumber
\end{eqnarray}
for the covariant derivatives will show that the background is
on-shell. Here $A$, $B$ and $C$ corresponds to any tangent space
index. At dimension $3/2$ we have
\begin{eqnarray}
\begin{array}{lll}
\label{FSa}
[\nabla_\alpha , \nabla_{\beta\dot\beta}] = \varepsilon_{\alpha \beta} \bar W_{\dot \beta},
    &[\nabla_\alpha, \bar \nabla_\il\, ]=\bar F_{\alpha \il},
    &[\bar \nabla_{\dot \alpha} , \bar \nabla_\il ]=0,
\end{array}
\end{eqnarray}
together with their complex conjugates. Note that the first and last
equations are a consequence of the algebra (\ref{algebrasolution})
plus Jacobi identities, while the second is the definition of $\bar
F_{\alpha\il}$. To proceed, we have to solve order-by-order in
$\theta$s using the Jacobi identities. Four-dimensional Lorentz
invariance implies that the first components of the superfields
defined above vanish and their second components should be
four-dimensional scalars, as discussed above. The field-strengths
$(W_\alpha, \bar F_{\alpha \il})$ have an expansion in powers of
$\theta$s. In particular we have the components
\ba\label{firstorder} W_\alpha = \theta_\alpha \mathrm D
+\theta^\il_\alpha h_\il\, + ...  \quad\quad \bar F_{\alpha \il}=
\theta_\alpha \bar
{\mathrm F}_\il\, +\theta^\jl_\alpha R_{\bar\il\jl}\, + ... \ea where 
the ellipses denote
components that do not concern us at the moment. The background
defined by (\ref{algebrasolution}) will be $N=1$ supersymmetric if
and only if these components vanish since all field-strengths should
be invariant under shifts of
$(\theta^\alpha,\bar\theta^{\dot\alpha})$. This is related to the
usual $N=1$ field theory requirement that in order to have a
supersymmetric vacuum, $D$ and $F$ terms should vanish. The
$h_\il\,$ and $R_{\bar\il\jl}\,$ components are, at this stage, not
required to vanish and are related to the geometry of the
compactified space. We will now calculate the values of these
components in terms of higher-dimension field-strengths.

 Using the Jacobi identities again we can alternative forms
 of the field-strengths:
\begin{eqnarray}\label{alternate}
\begin{array}{lll}
\label{FSb}
[\nabla_{\alpha \il} , \nabla_{\beta\dot\beta}] = \varepsilon_{\alpha \beta}  F_{\dot \beta \il}\, ,
        &[\nabla_{\alpha \il}, \nabla_\jl]= -2\epsilon_{\il\jl\kl} \bar F_{\alpha\kl}\, ,
    &[\nabla_{\alpha \il}, \bar \nabla_\jl]=-\delta_{\il\bar\jl} W_\alpha.
\end{array}
\end{eqnarray}
At lowest order in $\theta$ the $\bar {\mathrm F}_\il$ component
inside $\bar F_{\alpha\il}$ is given by $\{ \nabla_\alpha , \bar
F_{\beta\il} \} =\varepsilon_{\alpha\beta} \bar {\mathrm F}_\il$.
However, using (\ref{alternate}) we can write $\bar F_{\alpha\il}$
as \be \bar F_{\alpha\il} = \frac{1}{2} \epsilon_{\il\jl\kl}
[\nabla_{\alpha\jl},\nabla_\kl].\ee Now, the $ \{\nabla_\alpha ,  [
\nabla_{\beta\jl} , \nabla_{\kl}] \}$ Jacobi identity implies that
\be \bar {\mathrm F}_\il = \frac{i}{2} \epsilon_{\il\jl\kl} [
\nabla_\jl\, , \nabla_\kl\, ]\ee and since $[ \nabla_\jl\,
,\nabla_\kl\, ]$ is anti-symmetric, it follows that the vanishing of
the component $\bar {\mathrm F}_\il$ implies that $[ \nabla_\jl\,
,\nabla_\kl\, ]=0$. As we saw in section 2, these two conditions
imply that the compactification manifold is K\"ahler and that the
vector bundle over it is holomorphic.

In a similar way, the component ${\mathrm D}$ of $W_\alpha$ is the
lowest component of $ \{ \nabla_\alpha , W_\beta \} =
\varepsilon_{\alpha\beta} {\mathrm D}$. The computation of its value
in terms of higher dimension field-strengths has one additional
step. First we have to use the Jacobi identity with $
\{\nabla_\alpha , [ \nabla_{\beta\il} , \bar\nabla_\jl\, ]\}$ to
find \be\label{mfbi} \delta_{\il\bar\jl}\{\nabla_\alpha, W^\alpha\}
= \{\nabla_{\alpha\il} ,\bar F^{\alpha}_\jl\} +4i [\nabla_\il\,
,\bar\nabla_{\jl} ]\ee
Next, we use the Jacobi identity with $\{\nabla_{\alpha\il}
,[\nabla_{\beta_\jl}, \nabla_\kl\, ]\}$ to find \be \{
\nabla_{\alpha\il}\, , \bar F^{\alpha}_\jl \} = -4i [\nabla_\il\,
,\bar\nabla_\jl\, ] + 2i \delta_{\il\bar\jl}
\delta^{\kl\bar\ll}[\nabla_\kl\, , \bar\nabla_\ll\, ].\ee
Plugging this result back into (\ref{mfbi}) we find \be
\{\nabla_\alpha,W^\alpha\}= 2i \delta^{\kl\bar\ll}[\nabla_\kl\,
,\bar\nabla_\ll] .\ee
This means that the ${\mathrm D}$ component of $W_\alpha$ vanishes
when $\delta^{\kl\bar\ll}[\nabla_\kl\, ,\bar\nabla_\ll]=0$. This
equation is the second condition imposed by four-dimensional
supersymmetry.

In summary, we have found that the vanishing of $F$-terms in the
superfield $\bar F_{\alpha\il}$ implies the K\"ahler condition on
the compactified manifold and part of holomorphic YM equations for
the gauge background. The vanishing of the $D$-term in the
$W_\alpha$ field-strength implies Ricci-flatness and the remaining
equation for the set of holomorphic YM equations. One can proceed to
find the values of the other components of the field-strengths and
compute the expression for the curved $d$-operators in
(\ref{algebrasolution}) explicitly. For example, one can use the
Jacobi identity with $\{\nabla_{\alpha\il}\, , [\nabla_{\beta\jl}\,
, \nabla_\kl\, ]\}$ to find that $R_{\bar\il\jl} = -2i[\nabla_\jl\,
,\bar\nabla_\il\, ]$. The component $h_\il$ of $W_\alpha$ vanishes
due to the K\"ahler condition and the Jacobi identity with
$\{\nabla_{\alpha\il}\, ,[\nabla_{\beta\jl}\, , \bar\nabla_\kl\,
]\}$.

\section{Physical State Conditions and Spectrum}

Now that we have a BRST operator for the compactified background we
want to check that $Q(V)=0$ on a ghost  number one vertex operator
$V$ gives the correct spectrum for the compactification. In order to
do this we will first show that we get the correct equations of
motion for a super-Maxwell multiplet plus three chiral fields with
$N=1$ supersymmetry in four dimensions and then generalize to the
full string.

\subsection{ Ten dimensional super-YM in $1+3$ notation}

The propose of this section is to see how standard $N=1$ superfield
equations of motion appear when we perform a toroidal reduction of
the ten dimensional ghost number one vertex operator and BRST
charge. The vertex operator takes the form \be\label{flatvertex} V=
\lambda^\alpha A_\alpha +\bar\lambda^{\dot\alpha}\bar A_{\dot\alpha}
+\lambda^{\alpha\il} A_{\alpha\il} + \bar\lambda^{\dot\alpha
\il}\bar A_{\dot\alpha\il} \ee where $(A_\alpha, \bar
A_{\dot\alpha},A_{\alpha\il}\, ,\bar A_{\dot\alpha\il}\, )$ are
superfields of the full superspace. The solution of $QV=0$ where $Q$
is given by equation (\ref{flatq}) is
\be\label{flateqs}\begin{array}{l} d_\alpha A_\beta + d_\beta
A_\alpha =0 \cr d_\alpha A_{\beta\il} + d_{\beta\il} A_\alpha =
\varepsilon_{\alpha\beta} \bar\Phi_\il \cr d_{\alpha\il}A_{\beta\jl}
+d_{\beta\jl}A_{\alpha\il} =2\varepsilon_{\alpha\beta}
\epsilon_{\il\jl\kl}\, \Phi_{\kl} \cr d_\alpha \bar A_{\dot\alpha} +
\bar d_{\dot\alpha} A_\alpha =A_{\alpha\dot\alpha} \cr d_{\alpha\il}
\bar A_{\dot\alpha\jl} + \bar d_{\dot\alpha\jl} A_{\alpha\il}
=\delta_{\il\bar\jl\, } A_{\alpha\dot\alpha}\cr \bar
d_{\dot\alpha}\bar A_{\dot\beta} +\bar d_{\dot\beta}\bar
A_{\dot\alpha} =0 \cr \bar d_{\dot\alpha}\bar A_{\dot\beta\il} +
\bar d_{\dot\beta\il} \bar A_{\dot\alpha} =
\varepsilon_{\dot\alpha\dot\beta}\Phi_\il \cr \bar
d_{\dot\alpha\il}\bar A_{\dot\beta\jl} +\bar d_{\dot\beta\jl}\bar
A_{\dot\alpha\il} =2\varepsilon_{\dot\alpha\dot\beta}
\epsilon_{\il\jl\kl}\,\bar\Phi_{\kl} \cr d_\alpha \bar
A_{\dot\beta\il} + \bar d_{\dot\beta\il} A_\alpha = 0 \cr \bar
d_{\dot\alpha} A_{\beta\il} + d_{\beta\il} \bar A_{\dot\alpha} = 0,
\end{array}
\ee where the $d$s are defined in (\ref{newrealization}) and
$(A_{\alpha\dot\alpha},\Phi_\il\, ,\bar\Phi_\il\, )$ are defined by
these equations. The vertex operator $V$ also has the gauge
invariance $\delta V= Q\Lambda$ with a real superfield $\Lambda$. In
terms of its components, this translates to \be \delta
A_\alpha=d_\alpha \Lambda,\quad \delta A_{\alpha\il} =d_{\alpha\il}
\Lambda \ee together with their complex conjugates. The first
equation in (\ref{flateqs}) implies that $A_\alpha = d_\alpha V$ for
some complex superfield $V$. The equations of motion imply the
following gauge invariance \be \delta A_{\alpha\dot\alpha}
=\partial_{\alpha\dot\alpha} \Lambda,\quad \delta\Phi_\il =
\bar\partial_\il \Lambda,\quad \delta\bar\Phi_\il
=\partial_\il\Lambda \ee

We can use the algebra of the supersymmetric derivatives to derive
various relations on the fields defined by (\ref{flateqs}). It is
possible to solve all the Bianchi identities for a general set of
$(A_{\alpha\dot\alpha},\Phi_\il\, ,\bar\Phi_\il\, )$ but since our
goal is to generalize this to the case of a CY compactification, we
will take another route. First, note that it is possible to fix
$(A_\alpha,\bar A_{\dot\alpha})$ to vanish without trivializing the
system of equations. The gauge transformation that preserves this
choice has to satisfy \be d_\alpha\Lambda = \bar
d_{\dot\alpha}\Lambda =0,\ee which, by use of the $d$-operator
algebra, means $\Lambda$ is just a constant in four dimensions. This
implies that  the degrees of freedom described by $
(0,0,A_{\alpha\il}\, ,\bar A_{\dot\alpha\jl}\, )$ do not have gauge
invariance from the four-dimensional point of view.

When $A_\alpha=\bar A_{\dot\alpha}=0$ the equations (\ref{flateqs})
simplify to \be\begin{array}{ll} d_\alpha A_{\beta\il}
=\varepsilon_{\alpha\beta}\bar\Phi_\il, & \bar d_{\dot\alpha}
A_{\beta\il} =0\cr d_{\alpha\il} A_{\beta\jl} +d_{\beta\jl}
A_{\alpha\il} =2\varepsilon_{\alpha\beta}\epsilon_{\il\jl\kl}
\Phi_\kl, & d_{\alpha\il} \bar A_{\dot\alpha\jl} + \bar
d_{\dot\alpha\jl} A_{\alpha\il} =0 \cr \bar d_{\dot\alpha\il} \bar
A_{\dot\beta\jl} +\bar d_{\dot\beta\jl} \bar A_{\dot\alpha\il}
=2\varepsilon_{\dot\alpha\dot\beta}\epsilon_{\il\jl\kl} \bar\Phi_\kl
& \bar d_{\dot\alpha} \bar A_{\dot\beta\il}
=\varepsilon_{\dot\alpha\dot\beta} \Phi_\il
\end{array}
\ee

The first equation can be used to show that $d_\alpha
\bar\Phi_\il=0$, so it describes an anti-chiral field. The second
equation together with the first shows that $\bar d^2
\bar\Phi_\il=0$, which is the massless equation for a chiral field.
Then, using the commutator $[\bar d^2, \bar d_{\dot\alpha\il} ]=-
4i\bar d_{\dot\alpha}\bar\partial_\il$ and a combination of the
equations above, we find that \be \bar d^2\bar\Phi_\kl
\epsilon_{\kl\il\jl}=-2i(\bar\partial_\il\Phi_\jl -
\bar\partial_\jl\Phi_\il) =0, \ee which indicates that the massless
field equation of the chiral superfield is related to the cohomology
of 1-forms. If we set the higher
$(\theta^\il_\alpha,\bar\theta^\il_{\dot\alpha})$ components to
zero, we have precisely a triplet of chiral and anti-chiral fields.
We also need to determine the higher
$(\theta^\il_\alpha,\bar\theta^\il_{\dot\alpha})$ components. This
is accomplished by computing $d_{\alpha\il}\Phi_\jl$ and $\bar
d_{\dot\alpha\il}\Phi_\jl$. Using the equations above and the $d$
algebra, we find that \be d_{\alpha\il} \Phi_\jl =
-2i\bar\partial_{\jl} A_{\alpha\il},\quad\quad \bar
d_{\dot\alpha\il} \Phi_\jl = -2i \bar\partial_\jl \bar
A_{\dot\alpha\il}, \ee so the higher
$(\theta^\il_\alpha,\bar\theta^\il_{\dot\alpha})$ components do not
describe new degrees of freedom. If the fields do not depend on
$(z^\il\, , \bar z^\il\, )$ we have a triplet of four-dimensional
chiral fields, as desired.

It is easy to check that if we try to impose $A_{\alpha\il}=\bar
A_{\dot\alpha\il}=0$, we get a trivial system. Similarly, a solution
where $\Phi_\il=0$ and $\bar\Phi_\il=0$ is trivial because the
vector field strength $W_\alpha$ is a higher component in $\Phi$.
There is no covariant way to solve the constraints containing only
the gauge part. However, if the fields do not depend on the internal
coordinates, it is possible to isolate the four-dimensional gauge
part. Instead of following this path, it is worthwhile to derive the
equations of motion from (\ref{flateqs}) for a general
$(A_{\alpha\dot\alpha}=i[d_\alpha,\bar d_{\dot\alpha}]V,\Phi_\il\, ,
\bar\Phi_\il\, )$. Repeated application of the $d$-algebra gives
\be\label{delphiV} \bar d_{\dot\alpha} ( \Phi_\il +
2\bar\partial_\il V) =0,\quad d_\alpha (\bar\Phi_\il - 2\partial_\il
V)=0,\ee \be d^2\Phi_\il +2i\epsilon_{\il\jl\kl}
\partial_\jl\bar\Phi_\kl=2\bar\partial_\il d^\alpha A_{\alpha},\quad \bar d^2\bar\Phi_\il
+2i\epsilon_{\il\jl\kl} \bar\partial_\jl \Phi_\kl=2\partial_\il \bar
d^{\dot\alpha} \bar A_{\dot\alpha}, \ee \be d^\alpha \bar d^2
d_\alpha V -2 \delta^{\il\bar\jl}(\partial_\il \Phi_\jl -
\bar\partial_\jl \bar\Phi_\il)=0,\ee where higher components of
$(\theta^\il,\bar\theta^\jl)$ (which are consequences of the
equations above) are set to zero. These are the linearized equations
of motion for ten dimensional superYM in $1+3$ notation obtained
long ago in reference \cite{Marcus:1983wb}. If the fields do not depend on the
internal coordinates, we get three chiral fields and a vector
multiplet. The higher components are again determined by
equation (\ref{flateqs}).

\subsection{Heterotic string spectrum}

The spectrum of the heterotic string is calculated in a similar way
by repeated application of the curved space derivative algebra and
the equations of motions coming from $QA=0$. Additionally, we now have to 
remember that the covariant derivatives act appropriately on each
section of the various vector bundles over the CY. We will see that when the section of vertex
operator is not in the cohomology of $\nabla_\il$, the state
corresponds to a Kaluza-Klein mode obeying a massive superspace equation of
motion.

We begin with the compactification dependent sector. The complete
heterotic string vertex operator must be tensored with the
right-moving dimension $1$ currents\footnote{In order to get a
dimension $(0,0)$ vertex operator we should also multiply by the
right-moving ghost $\bar c$.} given by $(\bar\partial
x^a,\bar\partial y^\il,\bar\partial\bar y^\il, \bar\mathbf
J_\Sigma)$, where $\Sigma$ is a general index for the two $E_8$
algebras. Although not discussed in the present paper,\footnote{In
the pure spinor formalism this comes from conservation of the BRST
current and the anomaly in the conservation of ghost and gauge
currents.} the anomaly cancelation condition of the $B$-field should
be taken into account. The simplest way to solve it is by the
standard embedding. This embedding breaks
one of the $E_8$ factors into $E_8 \to E_6\times SU(3)$. The
Ka{c}-Moody currents are decomposed into \be \bar\mathbf J_\Sigma\to
(\bar\mathbf J_\sigma\, , \bar\mathbf J_\rho\, ,\bar\mathbf
J_A^\il\, , \bar\mathbf J_{\bar A}^{\bar\jl}\, , \bar\mathbf
J^{\il\bar\jl}\, )\ee where $\sigma$ is the index of the adjoint
representation of $E_8$, $\rho$ is an index for the adjoint
representation of $E_6$, $A$ is the index for the fundamental
representation of $E_6$, and $(\il\bar\jl)$ are indices for
endomorphisms of the
holomorphic tangent bundle.

The BRST charge is now \be\label{curvedbrst} Q=\oint(
\lambda^\alpha\nabla_\alpha+\bar\lambda^{\dot\alpha}\bar\nabla_{\dot\alpha}
+\lambda^{\alpha\il}\,\nabla_{\alpha\il}
+\bar\lambda^{\dot\alpha\il}\,\bar\nabla_{\alpha\il}\, ).\ee We
proceed exactly as in the previous section. The equations from the
BRST physical state condition are of the form (\ref{flateqs}) with
the operators $d$ replaced by the operators $\nabla$ of
(\ref{curvedbrst}). As in the previous section, we will set
$A^\Gamma_\alpha=\bar A^\Gamma_{\dot\alpha}=0$ where $\Gamma$ denotes any
right-moving index. After doing this, we obtain the equations
\be\begin{array}{ll} \nabla_\alpha A_{\beta\il}^\Gamma
=\varepsilon_{\alpha\beta}\bar\Phi_\il^\Gamma, & \bar \nabla_{\dot\alpha}
A_{\beta\il}^\Gamma =0\cr \nabla_{\alpha\il} A_{\beta\jl}^\Gamma
+\nabla_{\beta\jl} A_{\alpha\il}^\Gamma
=2\varepsilon_{\alpha\beta}\epsilon_{\il\jl\kl} \Phi_\kl^\Gamma, &
\nabla_{\alpha\il} \bar A_{\dot\alpha\jl}^\Gamma +
\bar\nabla_{\dot\alpha\jl} A_{\alpha\il}^\Gamma =0 \cr
\bar\nabla_{\dot\alpha\il} \bar A_{\dot\beta\jl}^\Gamma
+\bar\nabla_{\dot\beta\jl} \bar A_{\dot\alpha\il}^\Gamma
=2\varepsilon_{\dot\alpha\dot\beta}\epsilon_{\il\jl\kl} \bar\Phi_\kl^\Gamma
& \bar\nabla_{\dot\alpha} \bar A_{\dot\beta\il}^\Gamma
=\varepsilon_{\dot\alpha\dot\beta} \Phi_\il^\Gamma
\end{array}
\ee
Using these equations and the algebra (\ref{algebrasolution}) we
obtain $\nabla_{\alpha}\bar\Phi_\il^\Gamma=0$ and
$\bar\nabla^2\bar\Phi_\il^\Gamma=0.$\footnote{We define
$\nabla^2=\nabla^\alpha\nabla_\alpha$ and
$\bar\nabla^2=\bar\nabla_{\dot\alpha}\bar\nabla^{\dot\alpha}$.} Note
that the commutator
$[\bar\nabla^2,\nabla_{\dot\alpha\il}]=-4i\nabla_{\dot\alpha}\bar\nabla_\il$
still holds for the covariant derivatives. This implies that the
chiral fields $\Phi^\Gamma_\il$ satisfy \be \bar\nabla_\il
\Phi_\jl^\Gamma - \bar\nabla_\jl \Phi_\il^\Gamma=0.\ee Thus, for
each type of index $\Gamma$, the corresponding chiral field is in
the cohomology ring $H^{0,1}( \mathfrak{T})$, where $\mathfrak{T}$
is the vector space corresponding to the index $\Gamma$. This is the
expected result for the matter part of a CY compactification. The
analysis of the higher $\theta$-components proceeds as in the
previous section. In particular, the $\Phi^\Gamma$ do not describe
additional degrees of freedom at the massless level.

To derive the equations of motion for the
compactification-independent part, we have to solve the
generalization of equations (\ref{flateqs}) with covariant
derivatives without setting the superfields $A^\Gamma_\alpha$ and
$\bar A^\Gamma_{\dot\alpha}$ to zero. Again, we obtain the
generalization of (\ref{delphiV}):

\be\label{delphiVcurv}
\bar\nabla_{\dot\alpha} ( \Phi_\il^\Gamma + 2\bar\nabla_\il V^\Gamma) =0,\quad
\nabla_\alpha (\bar\Phi_\il^\Gamma - 2\nabla_\il V^\Gamma)=0,\ee \be\label{kk}
\nabla^2\Phi_\il^\Gamma +2i\epsilon_{\il\jl\kl}
\nabla_\jl\bar\Phi_\kl^\Gamma=2\bar\nabla_\il \nabla^\alpha
A_{\alpha}^\Gamma,\quad \bar\nabla^2\bar\Phi_\il^\Gamma +2i\epsilon_{\il\jl\kl}
\bar\nabla_\jl \Phi_\kl^\Gamma=2\nabla_\il \nabla^{\dot\alpha}
A_{\dot\alpha}^\Gamma, \ee \be \nabla^\alpha \bar \nabla^2 \nabla_\alpha V^\Gamma
-2 \delta^{\il\bar\jl}(\nabla_\il \Phi_\jl^\Gamma - \bar\nabla_\jl
\bar\Phi_\il^\Gamma)=0.\ee

If the fields do not depend on the compactification, the three
possible right moving indices are the four-dimensional vector index,
the adjoint $E_8$ index, and the adjoint $E_6$ index. This completes
the massless spectrum of the heterotic string in the CY background.
As a final remark, since the equations above do not impose that the
fields are harmonic forms on the CY (see equation \ref{kk}), they
also describe in superspace the KK spectrum of the compactification.

\section{Discussion}

In this paper, we began the study of superstring compactifications
using the pure spinor formalism. Although only $N=1$ supersymmetry
in four dimensions is preserved, the description of the BRST
operator and spectrum given here uses the full superspace inherited
from ten dimensions. We first considered some algebraic aspects of
the compactification, mainly the BRST operator and the spectrum. In
a second paper we will discuss further aspects, such as the
construction of the sigma model describing the dynamics of the
compactification and the anomaly cancelation condition, which comes
from the conservation of the BRST current. In this discussion the
$B$-field, which played no role in the present work, will be
included.

One interesting direction for future work could be to see how the
well known non-renormalization theorems of Calabi-Yau
compactifications arise in the supersymmetric description given
here. This will require knowledge of the zero-mode measure for
scattering amplitudes (which will be presented elsewhere). It is
possible that the non-renormalization is just a consequence of the
superspace integration arising from this measure.

A more important line of research is to generalize these results to
Type II strings, especially in the case of flux compactifications
(for a review see {\it e.g.} \cite{grana}). Most of the results in
the literature use only supergravity methods and little is known
about $\alpha'$ corrections and the spectrum. Even though it is
unlikely that a sigma model including all powers of $\theta$ can be
written explicitly, partial knowledge will already be enough to
address important questions pertaining to the form of the effective
action of the light modes in a flux compactification. We plan to
address flux compactifications of the pure spinor formalism in the
future.

\section*{Acknowledgments}

We would like to thank P.A. Grassi, L. Mazzucato and D. Sorokin for
useful discussions. WDL and BCV also thank KITP at Santa Barbara,
where part of this work was done, for their kind hospitality. OC
would like to thank Galileo Galilei Institute for Theoretical
Physics at Arcetri for their kind hospitality, where parts of this
work were done. WDL thanks UNAB for the warm hospitality, where this
work was started. This work is supported by FONDECYT grants 1061050
and 7080027, UNAB grants DI-03-08/R and AR-02-09/R, NSF grants PHY
0653342, DMS 0502267, PHY 05-51164.


\end{document}